\documentclass[prx,floatfix,superscriptaddress,longbibliography,notitlepage, twocolumn, 10.5pt]{revtex4-2}
\usepackage[utf8]{inputenc}
\usepackage{graphicx}
\usepackage[dvipsnames]{xcolor}
\usepackage{float}
\usepackage[citecolor=RawSienna, colorlinks=true, linkcolor=MidnightBlue, urlcolor=MidnightBlue, linktocpage=true]{hyperref}
\usepackage[T1]{fontenc}
\usepackage{bbm}
\usepackage{amsmath}
\usepackage{lmodern}
\usepackage{amsfonts}
\usepackage{physics}
\usepackage{mathtools}
\usepackage{tikz}
\usepackage{mathtools}
\usepackage{appendix}
\usetikzlibrary{arrows.meta}
\usepackage{dsfont}
\usepackage{subcaption}
\definecolor{myred}{RGB}{187, 85, 102}
\definecolor{myblue}{RGB}{0, 68, 136}
\definecolor{myyellow}{RGB}{221, 170, 51}
\definecolor{mygreen}{RGB}{34,136,51}
\tikzstyle{vertex}=[circle, fill=myred, minimum size=5pt, inner sep=0pt, text=white]
\tikzstyle{vertex2}=[circle, fill=myred, minimum size=20pt, inner sep=0pt, text=white]
\tikzstyle{vertex_mt}=[circle, fill=myred, minimum size=45pt, inner sep=0pt, text=white]
\tikzstyle{vertex_mt2}=[circle, fill=myred, minimum size=60pt, inner sep=0pt, text=white]
\tikzstyle{vertexb}=[circle, fill=myblue, minimum size=45pt, inner sep=0pt, text=white]
\tikzstyle{vertexb_b}=[circle,very thick, fill=myyellow, minimum size=45pt, inner sep=0pt, text=black]
\tikzstyle{vertexblank}=[circle, fill=white, minimum size=1pt, inner sep=0pt, text=myyellow]
\tikzstyle{vertex_small}=[circle, fill=myyellow, minimum size=10pt, inner sep=0pt, text=white]
\tikzstyle{edge} = [draw,ultra thick,-]
\tikzstyle{dedge} = [draw,ultra thick, dashed]

\begin{document}
\newtheorem{assumption}{Assumption}

\title{Gadgets for simulating a non-native $XX$ interaction in quantum annealing}
\author{Robert J.~Banks}
\affiliation{London Centre for Nanotechnology, University College London, Gower Street, WC1E 6BT, London, United Kingdom}
\author{Natasha Feinstein}
\affiliation{London Centre for Nanotechnology, University College London, Gower Street, WC1E 6BT, London, United Kingdom}
\author{Roopayan Ghosh}
\affiliation{Department of Physics and Astronomy, University College London, Gower Street, WC1E 6BT, London, United Kingdom}
\author{Sougato Bose}
\affiliation{Department of Physics and Astronomy, University College London, Gower Street, WC1E 6BT, London, United Kingdom}
\author{P.~A.~Warburton}
\email{p.warburton@ucl.ac.uk}
\affiliation{London Centre for Nanotechnology, University College London, Gower Street, WC1E 6BT, London, United Kingdom}
\date{\today}

\begin{abstract}
In certain scenarios, quantum annealing can be made more efficient by additional $XX$ interactions. It has been shown that the additional interactions can reduce the scaling of perturbative crossings. In traditional annealing devices these couplings do not exist natively. In this work, we develop two gadgets to achieve this:  a three-body gadget that requires a strong $ZZZ$ interaction; and a one-hot gadget that uses only local $X$ drives and two-body $ZZ$ interactions. The gadgets partition the Hilbert space to effectively generate a limited number of $XX$ interactions in the low-energy subspace. We numerically verify that the one-hot gadget can mitigate a perturbative crossing on a toy problem. These gadgets establish new pathways for implementing and exploiting $XX$ interactions, enabling faster and more robust quantum annealing.
\end{abstract}

\maketitle

\section{Introduction}
 Adiabatic quantum optimisation \cite{Far00}, as well as the closely related quantum annealing algorithm \cite{Kad98},  have been presented as quantum algorithms for solving combinatorial optimisation problems. In the typical formulation of quantum annealing, the combinatorial optimisation problem is encoded as an Ising Hamiltonian. To perform the protocol, the system is initialised in the ground state of a homogeneous transverse-field driver. The Hamiltonian is then interpolated from the transverse field to the Ising Hamiltonian. If this is done sufficiently slowly, the adiabatic theorem guarantees that the system will end up in the ground state of the Ising Hamiltonian \cite{Alb18}. The timescale required for adiabaticity is typically taken to be proportional to the inverse square of the minimum of the spectral gap (i.e. the difference between the instantaneous ground- and first-excited states). Often the spectral gap closes exponentially, leading to exponentially long run times \cite{Alb18, Hau20}.

In works such as \cite{Alb19,Seo12,Sek15,Tak20,Tak21,Cro14,gho24,Fei24,fei24_2, sal25} the effect of adding $XX$ terms, where $X$ is the Pauli $X$ matrix, to a quantum anneal has been investigated. It has been shown that, in some cases, $XX$ interactions can lead to an increase in performance for the problems considered. However, on most  quantum annealing hardware, there is no native $XX$ interaction between qubits. Instead the qubits are driven homogeneously, experiencing a $-X$ \footnote{I.e. a field that couples the up and down state of each individual spin, with strength -1.} field. This is referred to as transverse-field driving. 

In this work, we show how an effective $-XX$ \footnote{I.e. an interaction that couples two-qubit states of the same parity, with interaction energy -1.} interaction between a pair of qubits can be realised, using standard homogeneous transverse-field driving. To do this, we make use of imposing constraints in the computational basis. Combinatorial optimisation problems often come with constraints on the feasible solutions. These constraints can be enforced by a large energy penalty \cite{Luc14} or by carefully designing the dynamics \cite{Hen16, Hen16_2}. Constraints have also been used to realise quantum simulation in quantum annealing \cite{wer24,Abe21}, primarily through domain-wall encoding \cite{Cha19}. In this work, we explore the addition of constraints enforced by large energy penalties to impose the desired dynamics. We explore two gadgets based on constraints. One is based on encoding the parity of the two qubits. Encoding qubits through their parity has previously been explored in quantum annealing; it has been shown that it can result in all-to-all connectivity \cite{Lec15}. The second gadget uses a one-hot encoding. A one-hot encoding refers to all valid configurations having one spin up and the rest spin down. The one-hot gadget constrains the Hamming weight of four physical qubits to realise an effective $-XX$ interaction between two logical qubits. 

The gadgets resemble perturbative gadgets used to analyse the complexity of local Hamiltonians \cite{Jor08, Bra08}. These have seen little direct application in adiabatic quantum optimization, though they have seen some application in gate-based quantum optimization \cite{Cic24}. The aim of this paper is to create gadgets that effectively generate a desired two-body interaction from a given set of Ising-like interactions (possibly two-body only) and transverse-field driving. In contrast, perturbative gadgets are designed to map higher order interactions into two- \cite{Jor08,Bra08} (or possibly three- \cite{Cic24}) body interactions. 

The next section details quantum annealing and the gadgets used to provide effective $-XX$ interactions. In this work we restrict ourselves to non-overlapping constraints (i.e.\ constraints that share no physical qubits in common). This limits the number of $-XX$ interactions that can be achieved. Sec.~\ref{sec:num_xx} presents a numerical example on a toy problem where we validate the performance of the one-hot gadget with adiabatic quantum optimisation. We show how the one-hot gadget, using only $ZZ$ interactions, can mitigate a vanishing spectral gap. 

In this paper, the respective Pauli matrices are denoted by $X$, $Y$ and $Z$. The identity is denoted by $I$ and $\hbar$ has been set to one throughout. The numerical simulations made use of the Python package QuTip \cite{Qutip1, Qutip2}.

\section{The gadgets}
\label{sec:theory}
In quantum annealing the system is evolved under the Hamiltonian
\begin{equation}
    H(t)=A(t)H_d+B(t)H_p,
\end{equation}
where $H_p$ is the problem Hamiltonian that encodes the optimisation problem, which typically has the form:
\begin{equation}
    \label{eq:ising}
    H_p=\sum_{i,j}J_{i,j}Z_iZ_j+\sum_i h_i Z_i.
\end{equation}
The driver Hamiltonian, $H_d$ is typically the transverse-field driver given by:
\begin{equation}
    \label{eq:tf}
    H_d=-\sum_i X_i.
\end{equation}
In the rest of this paper we discuss how a $-XX$ interaction can be effectively engineered using only the interactions present in Eq.~\ref{eq:ising} and Eq.~\ref{eq:tf}. We also discuss the case where three body $ZZZ$ interactions are available. Higher order $Z$ terms have clear application in improving encoding of combinatorial optimisation problems \cite{Cha16, Lei16}. To this end, we describe two gadgets:
\begin{enumerate}
    \item A three-body gadget that makes use of $ZZZ$ interactions.
    \item A one-hot gadget that only uses two-body $ZZ$ interactions.
\end{enumerate} 
Both gadgets produce an effective $-XX$ interaction. Since the one-hot gadget involves only two-body interactions, it is the focus of Sec.~\ref{sec:num_xx}, where it is explored numerically. 

The theoretical motivation behind the three-body and one-hot gadgets is as follows:
\begin{enumerate}
    \item Add auxiliary qubits to increase the size of the Hilbert space.
    \item Add penalty terms to partition the Hilbert space into high- and low- energy subspaces.
    \item Write down an effective Hamiltonian for the low energy subspace.
\end{enumerate}
As we will demonstrate, this allows us to generate effective $-XX$ interactions between pairs of logical qubits, provided that the constraints introduced by the penalty terms share no physical qubits. One drawback of this technique is that there is a limit to the number of effective $-XX$ interactions that can be generated.

\subsection{Low-energy Hamiltonians}
To start with, we summarise how an effective low-energy Hamiltonian can be formulated through partitioning; the derivation can be found in \cite{Nee20}. Given a Hamiltonian $H$ of the form
\begin{equation}
    \label{eq:mat_block}
    H=\begin{pmatrix}
    H_{AA} & H_{AB}\\
    H_{BA} & H_{BB} 
    \end{pmatrix}
\end{equation}
satisfying the eigenvalue equation:
\begin{equation}
\begin{pmatrix}
H_{AA} & H_{AB}\\
H_{BA} & H_{BB} 
\end{pmatrix}
\begin{pmatrix}
C_A \\
C_B 
\end{pmatrix}
=
E
\begin{pmatrix}
C_A \\
C_B 
\end{pmatrix},
\end{equation}
an effective Hamiltonian for the state described by $C_A$ can be written down as follows:
\begin{equation}
    \label{eq:heff}
    H_\text{eff}(E)=H_{AA}-H_{AB}(H_{BB}-EI)^{-1}H_{BA}.
\end{equation}
From here, we make the following assumptions:
\begin{enumerate}
\item The high-energy subspace of the Hamiltonian, $H_{BB}$, is approximately diagonal (this assumption can be improved upon by carefully expanding $(H_{BB}-EI)^{-1}$). This is equivalent to the diagonal elements being much larger than the off diagonal elements.
\item The couplings, $H_{AB}$, are sufficiently weak compared to the energy gap between $H_{AA}$ and $H_{BB}$. This requirement is necessary to make the partitioning stable against small perturbations. 
\item The energy $E$ can be approximated with the typical energy of the low-energy subspace.
\end{enumerate}
The above assumptions allow us to write down effective low energy Hamiltonians. The rest of the section details the gadgets used in this paper.

\subsection{Three-body gadget}
In this section, we show how transverse-field driving can lead to an effective $-XX$ interaction. This gadget provides the intuition for the one-hot gadget described in the next section. We allow ourselves to use a three body $ZZZ$ interaction in this section. The aim is to show how an effective $-X_1X_2$ interaction can be realised between logical qubits `1' and `2'. First, an auxiliary qubit is added, indexed by `12'. A penalty is added to partition the Hilbert space. The penalty is given by $-C_p Z_1Z_2Z_{12}$, with $C_p>0$. This partitions the states, such that all the low energy states have an even number of ones (namely: $\ket{000}$, $\ket{011}$, $\ket{101}$, and $\ket{110}$) and are all a Hamming distance of two away from each other. The qubit `12' is measuring the parity of the first two qubits under this constraint. 

The intuition for this set-up is as follows:
\begin{enumerate}
    \item Qubit `12' encodes the parity of qubits `1' and `2' under the constraint.
    \item If no drive is applied to qubit `12', the parity of qubits `1' and `2' is fixed. If qubits `1' and `2' are then driven, single body spin-flips cannot occur, as this would result in the parity of the qubits changing.
    \item Given that qubits `1' and `2' are being driven, we expect dynamics to occur. The effective dynamics should conserve the parity of qubits `1' and `2'. Hence, the effective dynamics should be proportional to some combination of $XX$ and $YY$. Since the local driving in this case does not take into account any information about phase, the effective dynamics reduces to being proportional to $XX$.
\end{enumerate}
Under the same argument, driving only qubits `1' and `12' should realise an $X_1$ rotation and driving only qubits `2' and `12' should realise an $X_2$ rotation. By simultaneously driving all the qubits, we therefore expect some combination of local $X$ terms and an $X_1X_2$ term.

Having established the intuition behind this gadget, we now demonstrate that this intuition is correct. Each qubit in the gadget is subjected to a transverse-field drive. The physical Hamiltonian is given by:
\begin{equation}
    H_\text{Phys}^\text{3-body}=-d_1 X_1-d_2X_2-d_{12}X_{12}-C_p Z_1Z_2Z_{12},
\end{equation}
where $d_i$, $i=1,2,12$ is the strength of the local drive, $C_p>0$ and $C_p \gg \abs{d_i }$. Writing this in the structure of Eq.~\ref{eq:mat_block} gives:
\begin{multline}
    H_\text{Phys}^{3-body}=\\    
    \bordermatrix{ & \text{\small{\textbf{000, 011, 101, 110}}} & \text{\small{\textbf{001, 010, 100, 111}}}
    \cr
      & -C_p I& -d_{12}I-d_1X_1-d_2X_2 \cr
      & -d_{12}I-d_1X_1-d_2X_2 & C_p I  },
\end{multline}
where the line on top of the matrix in bold shows the order of states in each sector. Hence, in this case $H_{AA}=-C_p I$, $H_{BB}=C_p I$ and $H_{BB}$ is diagonal. Its eigenvalues are trivially given by $C_p$. Therefore, in this case
\begin{equation}
    (H_{BB}-EI)^{-1}=\frac{1}{C_p-E}I.
\end{equation}
The interaction between subspaces is given by $H_{AB}=H_{BA}=-d_{12}I-d_1X_1-d_2X_2$. Evaluating Eq.~\ref{eq:heff} gives:
\begin{multline}
    H_\text{eff}^\text{3-body}(E)=-C_pI-\frac{1}{C_p-E} (-d_{12}I-d_1X_1-d_2X_2)^2.
\end{multline}
The only thing that remains is to set $E$. The eigenvalues of $H_{AA}$ are $-C_p$. We assume that the driving terms act as small perturbations to the eigenvalues, such that $E$ can be approximated with $-C_p$. The resulting effective Hamiltonian is (excluding terms proportional to the identity):
\begin{equation}
    \label{eq:3beff}
    H_\text{eff}^\text{3-body}=-\frac{d_1 d_{12}}{C_p} X_1-\frac{d_2 d_{12}}{C_p}X_2-\frac{d_1 d_{2}}{C_p} X_1X_2.
\end{equation}
The result is an effective $-XX$ interaction between logical qubits `1' and `2'. A $+XX$ interaction cannot be introduced without changing the sign of one of the logical single body $X$ terms \footnote{This can be overcome by using a high-energy Hamiltonian (i.e. using only states that violate the penalty constraint) and inhomogeneous driving.}. Terms diagonal in the computational basis carry through from $H_\text{Phys}^\text{3-body}$ to $H_\text{eff}^\text{3-body}$, with minimal change and without changing the off-diagonal terms. This is on the assumption that the eigenvalues of $H_{AA}$ and $H_{BB}$ can still be well approximated as $-C_p$ and $C_p$ respectively. Therefore, the penalty term should be much larger than the other diagonal terms. 

To keep terms on the same energy scale, all terms diagonal in the computational basis (excluding the penalty) should be scaled by $1/C_p$ in the physical Hamiltonian. Hence, the physical Hamiltonian:
\begin{multline}
    H_\text{Phys}^\text{3-body}=-C_p Z_1Z_2Z_{12}-d_1X_1-d_2X_2-d_{12}X_{12}\\
    +\frac{1}{C_p} \left(h_1Z_1+h_2Z_2+J_{12}Z_{12}\right),
\end{multline}
results in the effective low-energy Hamiltonian:
\begin{multline}
    \label{eq:heff3bp}
    H_\text{eff}^\text{3-body}=\frac{1}{C_p} \left(-d_1 d_{12} X_1-d_2 d_{12} X_2-d_1 d_{2} X_1X_2 \right.\\
    \left. +h_1Z_1+h_2Z_2+J_{12}Z_{1}Z_{2}\right),
\end{multline}
where the constraint on qubit `12' has been exploited. The factor of $1/C_p$ in front of Eq.~\ref{eq:heff3bp} means that the timescale associated with the dynamics is slowed by this factor. This presents a trade-off between accuracy (requiring a large value of $C_p$) and
 run-time (which scales proportional to $C_p$).

The three-body gadget shows how the imposition of a constraint can lead to an effective $-XX$ interaction, using three physical qubits. However, it requires the three-body interaction to be much stronger than any two-body interaction or one-body field. The realisation of such a set-up is challenging. In the next section we therefore show how the use of the three-body interaction can be circumvented.

\subsection{One-hot gadget}

\begin{table}
\centering
\begin{tabular}{ c|c } 
  Logical state & Physical state\\ 
  \hline
  00 & 0001 \\ 
  01 & 0010 \\
  10 & 0100 \\ 
  11 & 1000 \\
\end{tabular}
\caption{A table showing the one-hot encoding between logical and physical state. Note that the left most qubit in the physical states corresponds to physical qubit `1'.}
\label{tab:oh}
\end{table}

Using the intuition from the previous example, we shall now implement an effective $-XX$ interaction using only two-body interactions. We consider a set-up with four physical qubits. In this case, the two logical qubits are then encoded using four physical qubits and a one-hot encoding. To make the notation clearer, logical terms are denoted with a tilde, while physical terms have no tilde. If there is no distinction between a physical and a logical qubit, then these are also denoted without a tilde. Table \ref{tab:oh} summarises the one-hot encoding used. The penalty term introduced is 
\begin{equation}
    H^\text{OH}_\text{Pen}=C_p\left(\sum_{i=1}^4 Z_i-2\right)^2,
\end{equation}
where $C_p$ is positive and needs to be large enough to enforce the constraint throughout the dynamics. This constraint fixes the total Hamming weight of the four physical qubits to be 1. As with the previous encoding, all low-energy physical states are separated by a Hamming distance of 2. 

Each physical qubit is subjected to a transverse-field, with the local drive denoted by $d_i$. To enforce the constraint on the Hamming weight, it is required that $C_p\gg \abs{d_i}$. The physical Hamiltonian is given by:
\begin{equation}
    H_\text{Phys}^\text{OH}=-\sum_{i=1}^4d_iX_i+C_p\left(\sum_{i=1}^4 Z_i-2\right)^2.
\end{equation}
In matrix form, this has the following structure:
\begin{equation}
    H_\text{Phys}^\text{OH}=
    \bordermatrix{ &\textbf{ 1} & \textbf{0 \& 2} & \textbf{3} & \textbf{4} \cr
      & 0I_{4 \times 4} & H_{1,0\, \& \,2}&0&0 \cr
      & H_{0\, \&\, 2,1} & 4C_pI_{7 \times 7}&H_{0\, \& \, 2,3}&0 \cr
      & 0 & H_{3,0\,\& \, 2}&16 C_pI_{4 \times 4}&H_{3,4} \cr
      & 0 & 0&H_{4,3}&32C_pI_{1 \times 1}},
\end{equation}
where the labels on top of the matrix in bold denote the Hamming weight of the states involved in the associated subspace. The label \textbf{0 \& 2} denotes the subspace spanned by states of Hamming weight zero and two. The couplings between subspaces with different Hamming weight $i$ and $j$ is denoted by $H_{i,j}$, and $I_{k\times k}$ denotes the identity matrix with dimensions $k$. Since the transverse-field couples states which are separated by a Hamming distance one, each subspace of fixed Hamming distance is diagonal. In this context, the terms in Eq.~\ref{eq:mat_block} are $H_{AA}=0I_{4\times4}$, $H_{BB}=4C_pI_{7\times7}$, $H_{AB}=H_{1, 0\, \& \, 2}$. Following the same process as the three-body gadget (with the same assumptions) gives the effective Hamiltonian:
\begin{multline}
    H_\text{eff}^\text{OH}=-\frac{1}{4C_p}(d_1d_3+d_2d_4) \tilde{X}_1-\frac{1}{4C_p}(d_1d_2+d_3d_4) \tilde{X}_2\\
    -\frac{1}{4C_p}(d_2d_3+d_1d_4) \tilde{X}_1\tilde{X}_2\\
    +\frac{1}{4C_p}(d_1d_3-d_2d_4) \tilde{X}_1\tilde{Z}_2+\frac{1}{4C_p}(d_1d_2-d_3d_4) \tilde{Z}_1\tilde{X}_2\\
    +\frac{1}{4C_p}(d_1d_4-d_2d_3) \tilde{Y}_1\tilde{Y}_2.
\end{multline}
While this is a more complicated interaction than the three-body gadget, setting $d_1=d_2=d_3=d_4$ gives:
\begin{equation}
    H_\text{eff}^\text{OH}=-\frac{d_1^2}{2C_p} \tilde{X}_1-\frac{d_1^2}{2C_p} \tilde{X}_2
    -\frac{d_1^2}{2C_p} \tilde{X}_1\tilde{X}_2.
\end{equation}
This gadget provides the required $-XX$ interaction, but it is less tuneable than the three-body gadget. Since each logical qubit is now encoded across four physical qubits, the logical $Z$ terms can be implemented as follows:
\begin{align}
    \tilde{I}&\equiv\frac{1}{2}\left(Z_1+Z_2+Z_3+Z_4\right)\\
    \tilde{Z}_1&\equiv\frac{1}{2}\left(Z_1+Z_2-Z_3-Z_4\right)\\
    \tilde{Z}_2&\equiv\frac{1}{2}\left(Z_1-Z_2+Z_3-Z_4\right)\\
    \tilde{Z}_1\tilde{Z}_2&\equiv\frac{1}{2}\left(-Z_1+Z_2+Z_3-Z_4\right),
\end{align}
where the equivalence reflects that they are only equivalent in the low energy subspace. The choice of physical Hamiltonians to implement a logical term is not unique. Terms diagonal in the computational basis remain unchanged between the physical and effective Hamiltonians, under the same assumptions as the three-body gadget. To make sure that all terms in the effective Hamiltonian have the same energy scale, these terms in the physical Hamiltonian should be scaled by $1/2C_p$.

To summarise the one-hot gadget, the physical Hamiltonian 
\begin{multline}
    \label{eq:ohphysp}
    H_\text{Phys}^\text{OH}=C_p\left(\sum_{i=1}^4 Z_i-2\right)^2-d\sum_{i=1}^4X_i\\
    +\frac{1}{4 C_p}\left[h_1\left(Z_1+Z_2-Z_3-Z_4\right)+h_2\left(Z_1-Z_2+Z_3-Z_4\right)\right.\\
    \left.+J_{12}\left(Z_1-Z_2+Z_3-Z_4\right)\right],
\end{multline}
implements the effective low energy Hamiltonian in the space spanned by \{$\ket{0001}$, $\ket{0010}$, $\ket{0100}$, $\ket{1000}$\}:
\begin{multline}
    \label{eq:oheffp}
    H_\text{eff}^\text{OH}=\frac{1}{2C_p}\left[-d^2\left( \tilde{X}_1+\tilde{X}_2+ \tilde{X}_1\tilde{X}_2\right)\right.\\
    \left.+h_1 \tilde{Z}_1+h_2\tilde{Z}_2+J_{12}\tilde{Z}_1\tilde{Z}_2\right].
\end{multline}
Similar to the previous gadget, the timescale associated with the effective dynamics is increased. For the one-hot gadget, the time scales in proportion to $2C_p$, where $C_p$ is large. Eq.~\ref{eq:ohphysp} consists of only two-body interactions, in Sec.~\ref{sec:num_xx} we explore this gadget applied to a toy optimisation problem example.

\section{A numerical study on effective $-XX$ interactions}
\label{sec:num_xx}

\subsection{The problem instance}
\begin{figure}
\centering
\begin{tikzpicture}[scale=1, auto,swap]

\draw[rounded corners, myyellow, ultra thick] (-3, 1) rectangle (-1, -7) {};

\draw[rounded corners, myyellow, ultra thick] (1, 1) rectangle (3, -9) {};

\draw[myblue, ultra thick] (-2,0) .. controls (-4,-1) .. (-2,-2);

\foreach \pos/\name in { {(-2,0)/1}, {(-2,-2)/2}, {(-2,-4)/3}}
        \node[vertex_mt] (\name) at \pos {$\frac{(1+\delta W)}{n_0}$};

\foreach \pos/\name in { {(2,0)/n_0+1}, {(2,-2)/n_0+2}, {(2,-4)/n_0+3},{(2,-6)/n_0+4}}
        \node[vertex_mt] (\name) at \pos {$\frac{1}{n_0+1}$};

  
\foreach \source/ \dest  in {1/n_0+1, 1/n_0+2, 1/n_0+3, 1/n_0+4, 2/n_0+1, 2/n_0+2, 2/n_0+3, 2/n_0+4, 3/n_0+1, 3/n_0+2, 3/n_0+3, 3/n_0+4}
        \path[edge] (\source) -- node {}(\dest);

\node[vertex] () at (-2,-5.5) {};
\node[vertex] () at (-2,-6.0) {};
\node[vertex] () at (-2,-6.5) {};

\node[vertex] () at (2,-7.5) {};
\node[vertex] () at (2,-8.0) {};
\node[vertex] () at (2,-8.5) {};

\node[myyellow] () at (-2,1.3) {\Large{$G_0, n_0$}};
\node[myyellow] () at (2,1.3) {\Large{$G_1, n_0+1$}};
\node[myblue] () at (-4.2,-1) {\Large{$-XX$}};

\end{tikzpicture}
\caption{The toy problem, consisting of a weighted maximum-independent-set problem on a bipartite graph. The independent set on the left, called $G_0$, has $n_0$ nodes. The independent set on the right, called $G_1$, has $n_0+1$ nodes. Each node in $G_0$ has weight $\left(1+\delta W\right)/n_0$. Each node in $G_1$ has weight $1/\left(n_0+1\right)$. The black lines correspond to both edges in the graph and antiferromagnetic couplings in the Ising formulation. The blue line shows where the $-XX$ interaction is added in Sec.~\ref{sec:num_xx}. }
\label{fig:toy_problem}
\end{figure}
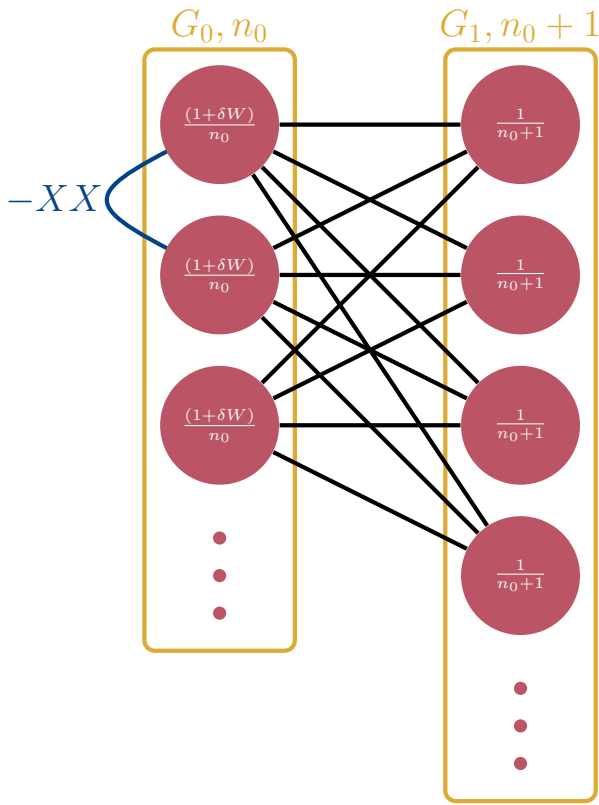

\begin{figure}
    \centering
    \includegraphics[width=0.48\textwidth]{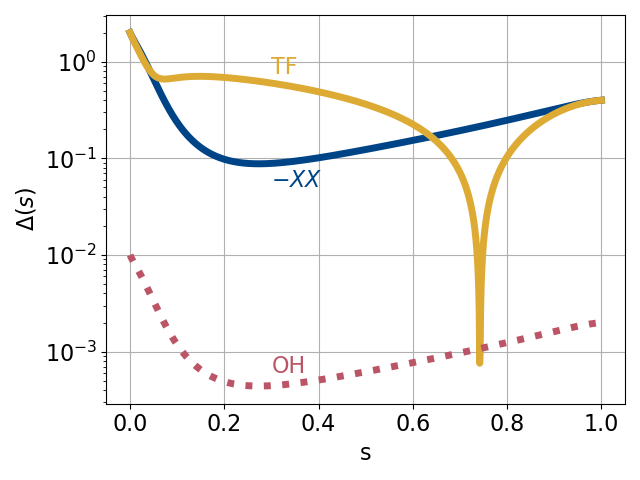}
    \caption{The spectral gap for the toy problem considered, with $n_0=10$, $J_{zz}=5.33$, $\delta W =0.1$, and $C_p=100$. The solid yellow line, marked `TF', shows the result for a standard transverse-field drive. The solid blue line, marked `$-XX$', shows the result when a $-XX$ interaction is introduced. The dashed red line, marked `OH', shows the result for the one-hot gadget. }
    \label{fig:spectral_gap_inst}
\end{figure}

In this section, we present a numerical study demonstrating the performance of the one-hot gadget detailed in Sec.~\ref{sec:theory}. We focus on this gadget as its implementation requires only two-body $ZZ$ interactions. The problem considered is an instance of a weighted maximum independent-set problem, detailed in \cite{Fei24}, on a bipartite graph. This problem is sketched out in Fig.~\ref{fig:toy_problem}. This toy problem has been chosen as it has been shown that $XX$ interactions can have a large impact on the success of adiabatic quantum optimisation on this problem, as well as it being relatively easy to scale. The problem Hamiltonian is given by:
\begin{multline}
    H_p=\sum_{i\in G_0, i=1}^{n_0}\left(n_1 J_{zz}-\frac{2\left(1+\delta W\right)}{n_0}\right) Z_i\\
    +\sum_{j\in G_1,j=n_0}^{2n_0+1}\left(n_0 J_{zz}-\frac{2}{n_1}\right) Z_j\\
    +J_{zz}\sum_{i\in G_0, j \in G_1 }Z_iZ_j
\end{multline}
where $G_0$ is an independent set with $n_0$ nodes and $G_1$ is an independent set with $n_0+1$ nodes. Throughout this paper, the parameters in the problem Hamiltonian are set to be $J_{zz}=5.33$ and $\delta W=0.1$. These parameters have been chosen as they generate a perturbative crossing \cite{Fei24, Ami09} at the end of the anneal, a known bottleneck in adiabatic quantum optimisation \cite{Alt10}. In this study, we focus on the spectral gap, $\Delta$, between the instantaneous ground- and first-excited states. The minimum spectral gap over the anneal is denoted by $\Delta_\text{min}$. The solid yellow line in Fig.~\ref{fig:spectral_gap_inst} shows the spectral gap for $n_0=10$, with a standard transverse-field driver
\begin{equation}
    H_d=-\sum_{i=1}^{2 n_0+1}X_i.
\end{equation}
The anneal is given by a linear interpolation:
\begin{equation}
    H(s)=(1-s)H_d+sH_p,
\end{equation}
where $s$ is the normalized time between $0$ and $1$. In the case of a standard transverse-field driver, there is a small gap towards the end of the spectrum.

The solid blue line in Fig.~\ref{fig:spectral_gap_inst} shows the spectral gap for the same linear interpolation when the driver is modified such that a $-XX$ interaction is added to the driver Hamiltonian between one pair of qubits in $G_0$. The modified driver Hamiltonian is 
\begin{equation}
    H_d^{(XX)}=-\left(\sum_{i=1}^{2 n_0+1}X_i\right)-X_1X_2,
\end{equation}
where $\{1,2\}\in G_0$, see Fig~\ref{fig:toy_problem}. The schedule for $H_d^{(XX)}$ is the same as the transverse-field case, i.e. $1-s$. The effect of the $-XX$ interaction is to soften the closing gap. In the next section, we show how this softening of the gap can be achieved with the one-hot gadget.

\begin{figure*}
\centering
\begin{tikzpicture}[scale=1, auto,swap]

\draw[rounded corners, mygreen, ultra thick] (-7, 1) rectangle (-3, -11) {};

\draw[rounded corners, mygreen, ultra thick] (3, 1) rectangle (5, -19) {};

\node[vertexb] (1h) at (-4,0) {$Z_{1h}$};
\node[vertexb] (4h) at (-4,-2) {$Z_{4h}$};
\node[vertexb] (2h) at (-6,-0) {$Z_{2h}$};
\node[vertexb] (3h) at (-6,-2) {$Z_{3h}$};
\foreach \source/ \dest  in {1h/2h, 1h/3h, 1h/4h, 2h/3h, 2h/4h, 3h/4h}
        \path[edge, myblue] (\source) -- node {}(\dest);

\foreach \pos/\name in {  {(-4,-4)/3}, {(-4,-8)/4}}
        \node[vertex_mt] (\name) at \pos {$Z_\name$};

\foreach \pos/\name in { {(4,0)/5}, {(4,-4)/6}, {(4,-8)/7},{(4,-12)/8},{(4,-16)/9}}
        \node[vertex_mt] (\name) at \pos {$Z_\name$};

  
\foreach \source/ \dest  in {1h/5, 1h/6, 1h/7, 1h/8, 1h/9, 4h/5, 4h/6, 4h/7, 4h/8, 4h/9,3/5, 3/6, 3/7, 3/8, 3/9,4/5, 4/6, 4/7, 4/8, 4/9}
        \path[edge] (\source) -- node {}(\dest);

\node[vertexb_b] (32) at (-4,-6) {$Z_{3}^{(2)}$};
\node[vertexb_b] (42) at (-4,-10) {$Z_{4}^{(2)}$};
\node[vertexb_b] (52) at (4,-2) {$Z_{5}^{(2)}$};
\node[vertexb_b] (62) at (4,-6) {$Z_{6}^{(2)}$};
\node[vertexb_b] (72) at (4,-10) {$Z_{7}^{(2)}$};
\node[vertexb_b] (82) at (4,-14) {$Z_{8}^{(2)}$};
\node[vertexb_b] (92) at (4,-18) {$Z_{9}^{(2)}$};

\foreach \source/ \dest  in {3/32, 4/42, 5/52, 6/62, 7/72, 8/82, 9/92}
        \path[edge, myyellow, ultra thick] (\source) -- node {}(\dest);

\foreach \source/ \dest  in {32/5, 32/6, 32/7, 32/8, 32/9,32/52, 32/62, 32/72, 32/82, 32/92, 42/5, 42/6, 42/7, 42/8, 42/9,42/52, 42/62, 42/72, 42/82, 42/92, 1h/52, 1h/62, 1h/72, 1h/82, 1h/92,4h/52, 4h/62, 4h/72, 4h/82, 4h/92, 3/52, 3/62, 3/72, 3/82, 3/92, 3/52, 3/62, 3/72, 3/82, 3/92, 4/52, 4/62, 4/72, 4/82, 4/92, 4/52, 4/62, 4/72, 4/82, 4/92}
        \path[edge, myyellow, dashed] (\source) -- node {}(\dest);

\node[mygreen] () at (-5,1.3) {\Large{$G_0, n_0=4$}};
\node[mygreen] () at (4,1.3) {\Large{$G_1, n_0+1=5$}};

\end{tikzpicture}
\caption{The implementation of the one-hot gadget and homogeneous driving for the toy problem with $n_0=4$. For the case with inhomogeneous driving, discussed and analysed in Sec.~\ref{sec:impOH}, everything in yellow can be ignored. The blue nodes show the nodes involved in the one-hot gadget. The solid blue lines show the presence of strong time-independent couplings, of strength $2C_p$, used to enforce the constraint on Hamming weight for the four blue nodes. The black lines are couplings used to implement the problem Hamiltonian and are time-varying couplers proportional to $J_{zz}/C_p$. The yellow nodes and edges show the required additions to recover homogeneous driving, discussed in Sec.~\ref{sec:hom}. In this case, the red nodes correspond to $Z_i^{(1)}$ and the yellow nodes $Z_i^{(2)}$. The solid lines show strong ferromagnetic couplings
equal to $-2C_p$. The dashed yellow lines are couplings used to implement the problem Hamiltonian and are time-varying couplers proportional to $J_{zz}/C_p$. Single-body fields are not shown in this figure.}
\label{fig:ohsketch}
\end{figure*}
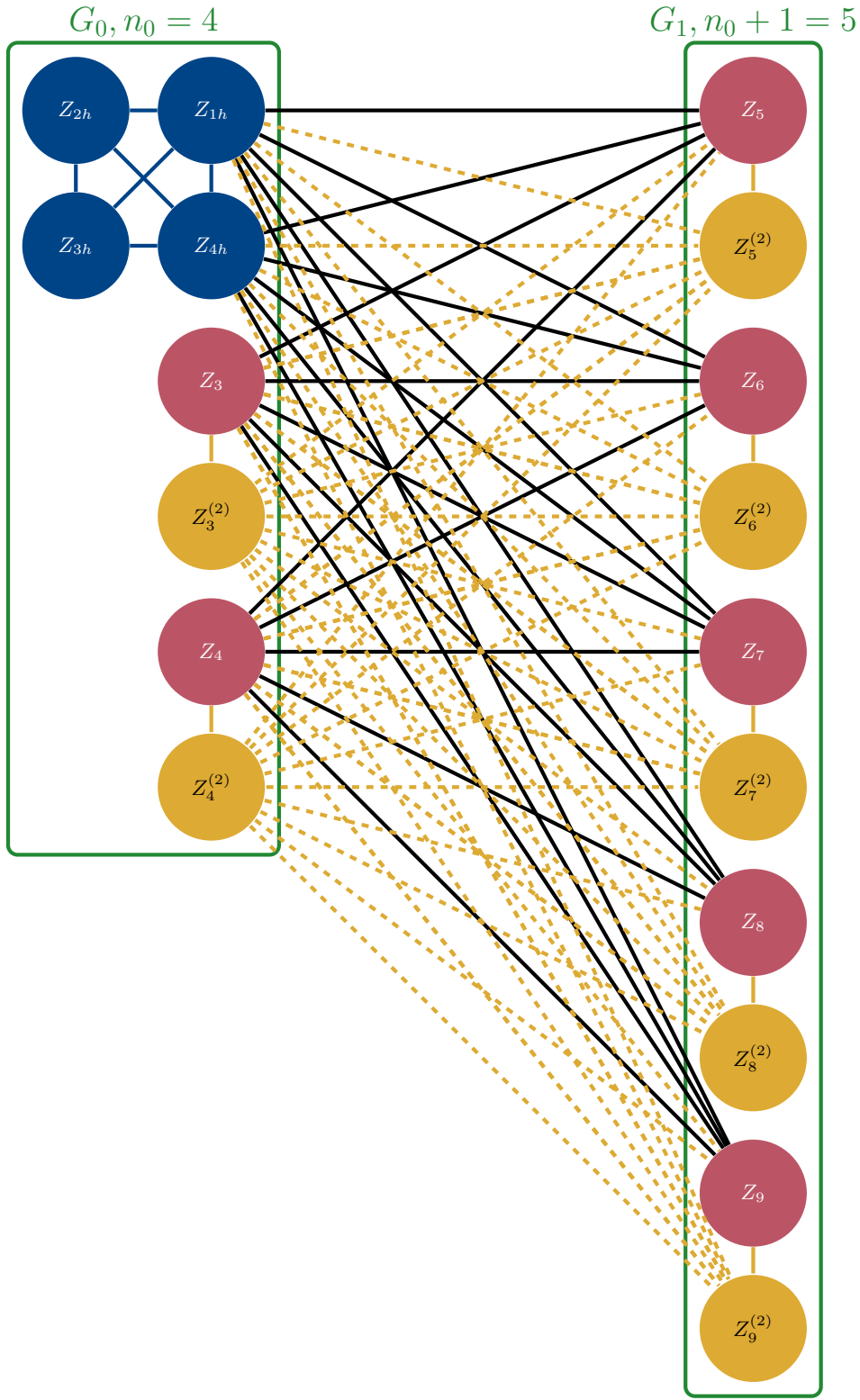

\subsection{Implementing the one-hot gadget}
\label{sec:impOH}
In this section, we show that the effect seen with the $-XX$ interaction can be achieved using the one-hot gadget introduced in Sec.~\ref{sec:theory}. 

For the one-hot gadget, the physical Hamiltonian to be implemented is given by: 
\begin{multline}
    \label{eq:OH_ham}
    H^{\text{OH}}(s)=C_p\left(\sum_{i=1}^4Z_{ih}-2\right)^2+\frac{(1-s)}{2 C_p}\left(-\sum_{i=3}^{2n_0+1}X_i\right)\\
    +\sqrt{1-s}\left(-\sum_{i=1}^4X_{ih}\right)
    +\frac{s}{2 C_p}H^{\text{OH}}_p,
\end{multline}
where
\begin{multline}
    H^{\text{OH}}_p=\sum_{i\in G_0, i=3}^{n_0}\left(n_1 J_{zz}-\frac{2\left(1+\delta W\right)}{n_0}\right) Z_i\\
    +\sum_{j\in G_1,j=n_0}^{2n_0+1}\left(n_0 J_{zz}-\frac{2}{n_1}\right) Z_j\\
    +J_{zz}\sum_{i\in G_0, j \in G_1, i\geq3}Z_iZ_j\\
    +\left(n_1 J_{zz}-\frac{2\left(1+\delta W\right)}{n_0}\right) \left(Z_{1h}-Z_{4h}\right)\\
    +J_{zz}\sum_{j \in G_1}Z_j\left(Z_{1h}-Z_{4h}\right),
\end{multline}
and the index `$h$' denoting qubits used in the one-hot gadget. Fig.~\ref{fig:ohsketch} sketches out the interactions required for the one-hot gadget on this problem with $n_0=4$. The $\sqrt{1-s}$ schedule on the driver Hamiltonian in the one-hot gadget comes from the effective Hamiltonian being proportional to the physical drive squared, as shown between Eq.~\ref{eq:ohphysp} and Eq.~\ref{eq:oheffp}. In Sec.~\ref{sec:hom} we discuss how homogeneous driving can be recovered. For the numerical simulations $C_p$=100. The dashed red line in Fig.~\ref{fig:spectral_gap_inst} shows the result for the spectral gap with $n_0=10$ with the one-hot gadget applied. Although the energy scale has changed, the gadget has given the correct shape compared to the $-XX$ modified driver.

\begin{figure}
    \centering
    \includegraphics[width=0.48\textwidth]{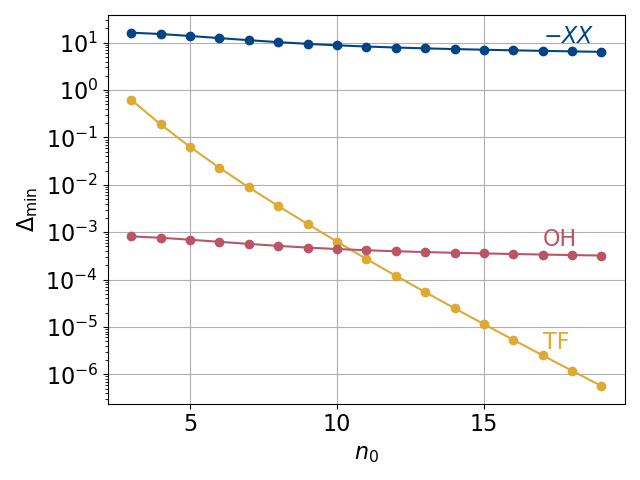}
    \caption{The scaling of the spectral gap with $n_0$ for the toy problem considered with $J_{zz}=5.33$, $\delta W =0.1$, and $C_p=100$. The solid yellow line, marked `TF', shows the result for a standard transverse-field drive. The  blue line, marked `$-XX$', shows the result when a $-XX$ interaction is introduced. The red line, marked `OH', shows the result for the one-hot gadget.}
    \label{fig:gap_scaling}
\end{figure}

The scaling of the spectral gap with $n_0$ is shown in Fig.~\ref{fig:gap_scaling}. In all instances $C_p=100$. Since the one-hot gadget has access to terms on the order of $C_p$, the non-gadget Hamiltonians have been multiplied by $C_p$ for fairer comparison. The standard transverse-field driver (yellow line) scales much worse than the true $-XX$ interaction (blue data) and the one-hot gadget (red data). The one-hot gadget achieves the same scaling as the $-XX$ case which it is emulating, despite the absolute value of the gap being much smaller. As the one-hot gadget scales better than the standard driver, at about $n_0=11$ it becomes favourable to use the gadget despite its energy scale being much smaller.

\begin{figure}
    \centering
    \includegraphics[width=0.48\textwidth]{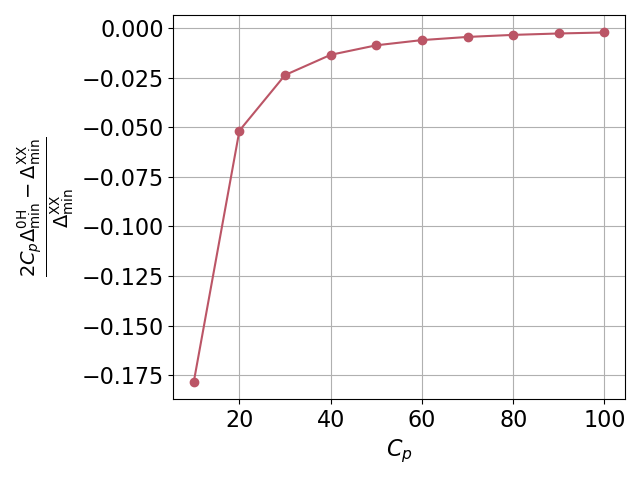}
    \caption{The error in the spectral gap between the direct implementation of $-XX$ and the one-hot gadget as a function of $C_p$ for $n_0$=10. The spectral gap for the one-hot gadget has been scaled by $2C_p$ for each data point. The line is a guide to the eye.}
    \label{fig:err_cp}
\end{figure}

Finally, to conclude this section, Fig.~\ref{fig:err_cp} shows the difference between the minimum spectral gap with a $-XX$ interaction ($\Delta_\text{min}^{XX}$) and the  minimum spectral gap with a one-hot gadget ($\Delta_\text{min}^\text{OH}$), as $C_p$ is swept. The spectral gap for the one-hot gadget has been scaled by $2C_p$ so the desired effective Hamiltonian matches the true $-XX$ interaction in energy scale. The difference is normalised by $\Delta_\text{min}^{XX}$. As is clear from the figure, the approximation breaks down as $C_p$ is decreased.

We have numerically shown in this section how the one-hot gadget, despite the addition of physical qubits and static two-body $ZZ$ terms, can improve the scaling of the spectral gap by successfully emulating a $-XX$ interaction. More generally, provided that the introduction of (non-overlapping) $-XX$ interactions improves the scaling of the minimum spectral gap, then there exists a critical size where the one-hot gadget will outperform the standard transverse-field driving for fixed $C_p$.

\subsection{Verifying dynamics}

\begin{figure}
    \centering
    \includegraphics[width=0.48\textwidth]{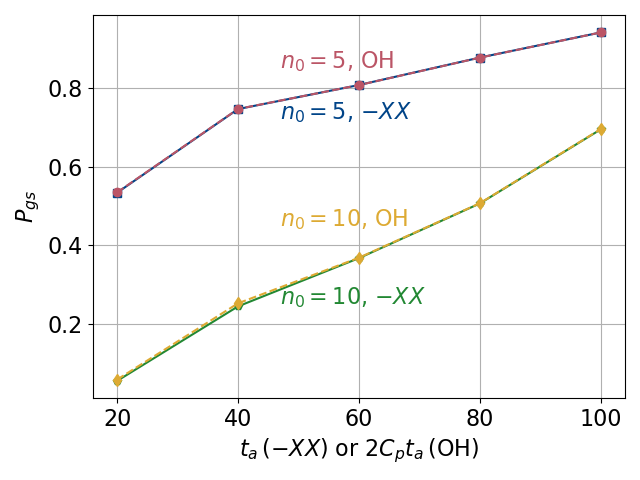}
    \caption{The ground-state probability $P_{gs}$ according to the Schr\"odinger equation. The anneal time for the $-XX$ interaction is denoted by $t_a$. The anneal time for the one-hot gadget is given by $2C_p t_a$, with $C_p=100$. The x-axis shows $t_a$. The red dots show the one-hot gadget with $n_0=5$. The blue dots show the $-XX$ interaction with $n_0=5$. The yellow dots show the one-hot gadget with $n_0=10$. The green dots show the $-XX$ interaction with $n_0=10$. The lines are guides to the eye.}
    \label{fig:gsp}
\end{figure}

In Fig.~\ref{fig:gsp} we plot the ground-state probability ($P_{gs}$) according to the Schr\"odinger equation, for both the one-hot gadget (Eq.~\ref{eq:OH_ham}) and directly implementing a $-XX$ interaction using linear schedules. The problem sizes considered are $n_0=5$ (with the one-hot gadget in red and the $-XX$ interaction in blue) and $n_0=10$ (with the one-hot gadget in yellow and the $-XX$ interaction in green). The anneal time for the $-XX$ interaction as $t_a$ is shown on the x-axis. The anneal time for the one-hot gadget is $2C_pt_a$. Since, $P_{gs}\not \approx 1$, the dynamics in this regime are not adiabatic. The one-hot gadget successfully emulates the Schr\"odinger evolution of the $-XX$ interaction, as desired and expected.

\subsection{Recovering homogeneous driving}
\label{sec:hom}
In the previous sections the physical qubits were inhomogeneously driven, i.e. the constrained qubits used to implement the one-hot gadget were subject to a different drive compared to the other qubits. To recover homogeneous driving between all physical qubits, a constraint can be applied to all the other physical qubits. This can be done by replacing each physical qubit not involved in the one-hot gadget with two physical qubits ferromagnetically chained together, with interaction strength $2 C_p$. This has the effect of reducing the effective drive on each logical qubit. The constraint applied fixes the parity of the two physical qubits:
\begin{equation}
    -2 C_p Z^{(1)}_iZ^{(2)}_i.
\end{equation}
The physical Hamiltonian to be implemented is
\begin{equation}
    H_\text{Phys}^\text{hom}=-d_1X^{(1)}_i-d_2X^{(2)}_i-2C_pZ^{(1)}_iZ^{(2)}_i,
\end{equation}
the resulting low-energy effective Hamiltonian is:
\begin{equation}
    H_\text{eff}^\text{hom}=-\frac{d_1d_2}{2 C_p} \tilde{X}_i.
\end{equation}
Hence the resulting drive can be mode homogeneous. Each logical $\tilde{Z}_i$ is encoded as two physical qubits by:
\begin{equation}
    \tilde{Z}_i\equiv\frac{1}{2}\left(Z^{(1)}_i+Z^{(2)}_i\right).
\end{equation}

\begin{figure}
    \centering
    \includegraphics[width=0.48\textwidth]{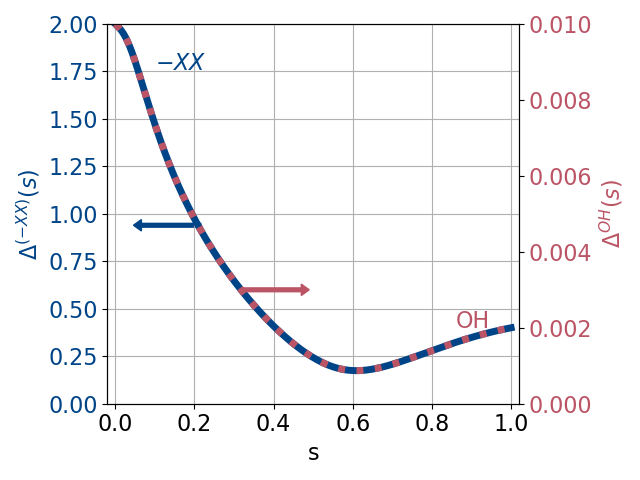}
    \caption{The spectral gap with homogeneous driving with $n_0=2$. The solid blue line shows the direct implementation of  the $-XX$ interaction. The dashed red line shows the spectral gap implementing both the one-hot gadget and inhomogeneous driving. The axis on the left is for  the`$-XX$' data, while the axis on the right is for the `OH' data.}
    \label{fig:hom}
\end{figure}

For the toy problem considered in Sec.~\ref{sec:impOH}, the Hamiltonian with homogeneous driving is given by:
\begin{multline}
    H^{\text{OH}}_\text{hom}(s)=C_p\left(\sum_{i=1}^4Z_{ih}-2\right)^2-2C_p \sum_{i=3}^{2n_0+1} Z_i^{(1)}Z_i^{(2)}\\
    +\sqrt{1-s}\left(-\sum_{i=3}^{2n_0+1}\left(X_i^{(1)}+X_i^{(2)}\right)-\sum_{i=1}^4X_{ih}\right)\\
    +\frac{s}{2 C_p}H^{\text{OH}}_{p,\text{hom}},
\end{multline}
with
\begin{multline}
    H^{\text{OH}}_{p,\text{hom}}=\\
    \sum_{i\in G_0, i=3}^{n_0}\frac{1}{2}\left(n_1 J_{zz}-\frac{2\left(1+\delta W\right)}{n_0}\right) \left(Z_i^{(1)}+Z_i^{(2)}\right)\\
    +\sum_{j\in G_1,j=n_0}^{2n_0+1}\frac{1}{2}\left(n_0 J_{zz}-\frac{2}{n_1}\right) \left(Z_j^{(1)}+Z_j^{(2)}\right)\\
    +\frac{J_{zz}}{4}\sum_{i\in G_0, j \in G_1, i\geq3}\left(Z_i^{(1)}+Z_i^{(2)}\right)\left(Z_j^{(1)}+Z_j^{(2)}\right)\\
    +\left(n_1 J_{zz}-\frac{2\left(1+\delta W\right)}{n_0}\right) \left(Z_{1h}-Z_{4h}\right)\\
    +\frac{J_{zz}}{2}\sum_{j \in G_1}\left(Z_j^{(1)}+Z_j^{(2)}\right)\left(Z_{1h}-Z_{4h}\right).
\end{multline}
Fig.~\ref{fig:ohsketch} sketches out the interactions required for the one-hot gadget on this problem with $n_0=4$ and homogeneous driving. The effect of introducing the one-hot gadget and homogeneous driving has been to double the number of physical qubits such that the number of physical qubits is twice the number of logical qubits. Fig.~\ref{fig:hom} shows the spectral gap for $n_0=2$ for the direct $-XX$ case (solid blue line) and the homogeneously driven case with the one-hot gadget (dashed red line). The spectral gap in the gadget case has been scaled by $2C_p$.

\subsection{Initial-state preparation}

Implicit in the work so far has been that the ground state of the effective Hamiltonian at $s=0$ can be reached, namely the ground-state of the effective driver Hamiltonian. This is not the conventional $\ket{+}$ state. We therefore discuss how to prepare the initial state in this section. The steps to prepare the ground state of the effective driver Hamiltonian can be achieved as follows:
\begin{enumerate}
    \item Select a physical state corresponding to a valid logical state according to the constraints (for example, one-hot constraints or three-body constraints). Let $z_i$ denote the initial state of the $i^\text{th}$ physical qubit (e.g. 0 or 1) of this valid configuration.
    \item Prepare the system in this initial state by initialising in the ground state of the Hamiltonian:
    \begin{equation}
    \label{eq:gs_prep}
       H_\text{init}=\sum_i (2z_i-1) Z_i.
    \end{equation}
    \item Turn on the terms, diagonal in the computational basis, in the Hamiltonian that enforce the constraint.
    \item Adiabatically interpolate between $H_\text{init}$ and $H_d=-\sum_iX_i$, with the constraints left on, to prepare the desired initial state at $s=0$.
\end{enumerate}

Since the constraints do not overlap, the run time for the state preparation will not scale with the problem size. This approach for state preparation closely resembles reverse quantum annealing \cite{Chancellor_2017}.

\section{Conclusion}
This work demonstrates how a $-XX$ interaction can be implemented in quantum annealing by applying constraints in the computational basis. The three-body gadget provided an intuitive way of realising a $-XX$ interaction. The one-hot gadget used this intuition, namely finding a set of four states mutually separated by a Hamming distance of two, to realise an effective $-XX$ interaction without three-body terms. With the addition of only static two-body $ZZ$ couplings, it was shown that the one-hot gadget could mitigate a perturbative crossing in adiabatic quantum optimisation. This work presents a first step towards realising desired Hamiltonians from an algorithmic perspective, reducing the need for new physical hardware interactions.

\section*{Acknowledgements}

This work was supported by EPSRC grant EP/Y004590/1 MACON-QC. The authors acknowledge the use of the UCL Myriad High Performance Computing Facility (Myriad@UCL), and associated support services, in the completion of this work.

\appendix

\bibliography{references.bib}

\end{document}